\documentclass[12pt]{article}
\usepackage{amsmath,amsfonts,amscd}
\usepackage{bm}
\usepackage{graphicx}
\def\cA{{\cal A}}
\def\cD{{\cal D}}

\def\cH{{\cal H}}             % Hilbert space
\def\cN{{\cal N}}

\def\cH{{\cal H}}

\newcommand{\R}{\mathbb{R}}
\def\Z{{Z\kern-.5em{Z}}}
\def\V{{V\kern-.7em{V}}}
\def\ie{{\em i.e.\, }}
\def\eg{{\em e.g.\, }}

\def\d{\partial}
\def\bra{\langle}
\def\ket{\rangle}

\def\lr{{\rm L}^2({\R^d})}
\def\bvec#1{{\bm #1}}      % bold vectors 
\def\vb{\bvec{b}}
\def\vk{\bvec{k}}
\def\vq{\bvec{q}}

\def\vx{\bvec{x}}
\def\dk#1#2{\frac{ d^{#2}{#1} }{ (2\pi)^{#2} }} % invariant measure in FT
\def\da#1#2{\frac{ d{#1}}{{#1}^{{#2}+1}}}
\begin{document}
\title{Wavelet based regularization for Euclidean field theory and 
stochastic quantization}
\author{M.V.Altaisky \\ 
\small Joint Institute for Nuclear Research, Dubna, 141980, Russia \\
\small and Space Research Institute, Moscow, 117997, Russia \\
\small e-mail: altaisky@mx.iki.rssi.ru}
\date{}
\maketitle
\begin{abstract}
It is shown that Euclidean field theory with polynomial interaction, can be regularized using the wavelet representation of the fields.
The connections between wavelet based regularization and stochastic 
quantization are considered with $\phi^3$ field theory taken as an example.
\end{abstract}
\section{Introduction}
The intimate connection between quantum field theory and stochastic
differential equations has called constant attention for
quite a long time \cite{Nelson1985,GJ1981}. 
Among different aspects of this connection 
the stochastic quantization, proposed by G.Parisi and Y.Wu \cite{PW1981}, 
is especially important, for it is believed to provide a new approach to 
quantization of Euclidean fields alternative to path integrals and 
canonical quantization. 
The stochastic quantization is also particularly attractive for 
quantization of gauge theories as it does not require gauge fixing. 
Besides that, the stochastic quantization procedure being based on 
random processes, and therefore on the measure theory, may be considered 
as a measure based formalism for constructing divergence free theory from 
beginning, instead of artificial regularization of the Riemann integrals in 
Euclidean space.

We know that stochastic processes often posses self-similarity. 
The renormalization procedure used in quantum field theory is also based on 
self-similarity. So, it is natural to use for the regularization of field
theories the wavelet transform (WT), the decomposition
with respect to the representation of the affine group.
In this paper two ways of wavelet based regularization are presented.
First, the direct substitution
of WT of the fields into the action functional leads to
a field theory with scale-dependent coupling constants. Second, the
WT, being substituted into the Parisi-Wu stochastic quantization
scheme  \cite{PW1981},
provides a stochastic regularization with no  extra vertexes introduced into the theory.

\section{Scalar field theory on affine group}
The scalar field theory with the polynomial interaction $V(\phi)$ 
defined on Euclidean space $\R^d$ is 
one of the most instructive models any textbook in field theory starts with, 
see \eg \cite{Ramond1981}. This field theory is defined by the generating 
functional 
\begin{equation}
W_E[J] = \cN \int \cD\phi \exp\left[-S_E[\phi(x)] + \int  d^dx J(x)\phi(x) \right] 
\label{gf},
\end{equation}
where $S_E[\phi(x)]$ is Euclidean action, $\cN$ is formal 
normalization constant. The (connected) Green  functions 
($m$-point cumulative moments) are evaluated 
as functional derivatives of the logarithm of generating functional 
$W_E[J] = e^{-Z_E[J]}$:
\begin{equation}
G_m(x_1,\ldots x_m) \equiv \bra \phi(x_1)\ldots\phi(x_m)\ket = -
\left. \frac{\delta^m}{\delta J(x_1)\ldots\delta J(x_m)}\right|_{J=0} 
\ln W_E[J].
\label{gfm}
\end{equation} 
The generating functional \eqref{gf} 
describes a quantum field $\phi(x)$  
with the action 
\begin{equation}
S_E[\phi] =\int_{\R^d} d^d x \left[{ 
\frac{1}{2}(\d_\mu\phi)^2+ \frac{m^2}{2}\phi^2 + \lambda V(\phi)}\right].
\label{fe}
\end{equation}
Alternatively, the theory of quantum 
field in Euclidean space is equivalent to the theory of classical fluctuating 
field with the Wiener probability measure 
$\cD P = e^{-S_E[\phi]}\cD\phi$. In this case 
$m^2 = |T-T_c|$ is the deviation from critical temperature and 
$\lambda$ is the fluctuation interaction strength. 

In the simplest case of a scalar field with the fourth power interaction 
\begin{equation}
S_E[\phi] =\int d^d x \left[{
\frac{1}{2}(\d_\mu\phi)^2+ \frac{m^2}{2}\phi^2 + \frac{\lambda}{4!}\phi^4
}\right]
\label{f4l}
\end{equation}
the theory is often referred to as the Ginsburg-Landau model for its ferromagnetic 
applications. 
The $\phi^3$ theory is also a useful toy model used, for instance, to describe 
electron-phonon interactions \cite{AGD}.

The straightforward way to calculate the Green functions is to factorize the 
interaction part  of generation functional \eqref{gf} in the form
\begin{equation}
W_E[J] = \exp\left[-V\left(\frac{\delta}{\delta J}\right) \right] 
\exp\left(-\frac{1}{2}JD^{-1}J \right),
\quad D = -\d^2 +m^2
\end{equation} 
The perturbation expansion is then evaluated in $k$-space, where 
$$
{\widehat D}^{-1}(k) = \frac{1}{k^2+m^2}.
$$
From  group theory point of view
the reformulation of a field theory from the coordinate representation 
$\phi(x)$ to the momentum representation $\hat\phi(k)$ by means  of 
Fourier transform 
$\hat\phi(k)=\int_{\R^d} e^{\imath k x} \phi(x) d^dx$, is only a 
particular case of decomposition 
of a function with respect to representation of a Lie group $G$. 
For the Fourier transform $G$ is just an abelian group of translations, 
but other groups 
may be used as well, depending on the physics of a particular problem.

For a locally compact Lie group $G$ acting transitively 
on the Hilbert space $\cH$  
it is possible to decompose vectors $\phi\in\cH$ with respect to the 
square-integrable representations $U(g)$ of the group $G$ 
\cite{Carey76,DM1976}:
\begin{equation}
|\phi\ket = C_\psi^{-1} \int_G |U(g)\psi\ket d\mu(g)\bra\psi U^*(g)|\phi\ket,
\label{pu}
\end{equation}
where $d\mu(g)$ is the left-invariant Haar measure on $G$.
The normalization constant $C_\psi$ is determined by the norm of the 
action of $U(g)$ on the fiducial vector $\psi\!\in\!\cH$, \ie any 
$\psi\!\in\!\cH$, that satisfies the admissibility condition 
\begin{equation}
C_\psi := \|\psi \|^{-2} \int_{g\in G} |\bra\psi|U(g)\psi \ket|^2 d\mu(g)
<\infty,
\label{adc}
\end{equation}
can be used as a basis of wavelet decomposition. 

In operator notation the decomposition \eqref{pu} is also referred to as 
a partition of unity with respect to the representation $U(g)$:
$$\hat1 = C_\psi^{-1} \int_G |U(g)\psi\ket d\mu(g)\bra\psi U^*(g)|.$$ 

It is apparently possible to generalize the  
Feynman-Dyson perturbation expansion in a given field theory $S_E(\phi)$ 
by using decomposition \eqref{pu} for non-abelian groups. 

For definiteness, let us consider the fourth power 
interaction model with the (Euclidean) action functional 
\begin{equation}
\begin{array}{lcl}
S[\phi] &=& \frac{1}{2}\int_{\R^d} \phi(x_1)D(x_1,x_2)\phi(x_2)dx_1dx_2  \\
        &+& \frac{\lambda}{4!}\int V(x_1,x_2,x_3,x_4)\phi(x_1)\phi(x_2)\phi(x_3)
          \phi(x_4)dx_1dx_2 dx_3dx_4,
\label{Sphi}
\end{array}
\end{equation}
where $D(x_1,x_2)$ is the inverse propagator of the free model ($\lambda\!=\!0$).
Using the notation 
$$U(g)|\psi\ket \equiv |g,\psi\ket, \quad \bra\phi|g,\psi\ket \equiv \phi(g),
\quad \bra g_1,\psi|D|g_2,\psi\ket \equiv D(g_1,g_2),
$$
we can rewrite the generating functional \eqref{gf} of the field theory 
with action \eqref{Sphi} in the form 
\begin{equation}
\begin{array}{lcl}
W_G[J(g)] &=& \int \cD\phi(g) \exp\Bigl(
-\frac{1}{2}\int_G \phi(g_1)D(g_1,g_2)\phi(g_2)d\mu(g_1) d\mu(g_2) \\
&-&\frac{\lambda_0}{4!}\int_G \tilde V(g_1,g_2,g_3,g_4)
\phi(g_1)\phi(g_2)\phi(g_3)\phi(g_4)
d\mu(g_1) d\mu(g_2)d\mu(g_3) d\mu(g_4)  \\
&+& \int_G J(g)\phi(g)d\mu(g)
\Bigr),
\end{array}
\label{gfi4}
\end{equation}
where $$
\begin{array}{lcl}\tilde V(g_1,g_2,g_3,g_4)&=&\int_{\R^d}V(x_1,x_2,x_3,x_4) 
(U(g_1)\psi(x_1))(U(g_2)\psi(x_2)) \\ 
& & (U(g_3)\psi(x_3))(U(g_4)\psi(x_4))dx_1dx_2dx_3dx_4
\end{array}
$$ 
is the result of application of the wavelet transform in all arguments 
of $V$. 
%$$\tilde\phi(g) := \int\overline{U(g)\psi(x)} \phi(x)dx$$ 
%to $V(x_1,x_2,x_3,x_4)$ in all arguments $x_1,x_2,x_3,x_4$.

At this point we ought mention a real-world problem arising from substitution of the functional integration over physical fields by 
integration over the fields defined on a Lie group $G$. For the abelian group 
of translations (homeomorphic to $\R^d$) the inverse transform from Euclidean 
space to Minkovski space by changing $t$ into $\imath\!t$ does not yield any 
problems with causality, since the chronological ordering of operators is 
easily imposed. For non-abelian groups it is not clear how to order, or how 
to commute, the operator-valued fields at different points of the Lie 
group manifold $[\phi(g_1),\phi(g_2)]=?$. This is an obstacle 
preventing quantization of gauge theories using wavelets
\cite{Federbush1995}. 
That is why we will further consider only $c$-valued fields, bearing in 
mind the interpretation of Euclidean field theory in terms of the 
models of statistical mechanics. 

Let us turn to the particular case of the affine group that is of our 
principle interest
$$
x'=aR(\theta)x+b,\quad R(\theta) \in SO_d, x,x',b \in \R^d
. 
$$
%%%%%%%%%%%%%%%%%%%%%%%%%%%%%%
Hereafter we assume the basic wavelet $\psi$ is invariant under 
$SO_d$ rotations $\psi(\mathbf{x})=\psi(|\mathbf{x}|)$ and drop
the angular part of the measure for simplicity.
After this simplifying assumption, the left-invariant 
Haar measure on affine group is  
$d\mu(a,\vb)=\frac{da d^d\vb}{a^{d+1}}$. The representation $U(g)$ induced 
by a basic wavelet $\psi(x)$  is 
\begin{equation}
g:\vx'=a\vx+\vb,\quad U(g)\psi(\vx)=a^{-d/2}\psi\left( \frac{\vx-\vb}{a}\right). 
\label{rag}
\end{equation}
The (bold) vector notation is dropped where it does not lead to confusion.
The last thing we need to construct the generating functional of a 
field theory on affine group is to substitute wavelet decomposition 
\begin{equation}
\phi(x)= C_\psi^{-1}\int a^{-d/2}\psi\left(\frac{x-b}{a}\right)\phi_a(b)
\frac{dad^db}{a^{d+1}},
\label{iwt}
\end{equation} 
into Euclidean action $S_E[\phi]$. Here we use normalization 
for the rotationally invariant wavelets
\begin{equation}
C_\psi = \int \frac{|\hat\psi(k)|^2}{S_d|k|^d}d^dk,
\label{adcf}
\end{equation}
where the area of the unit sphere in $d$ dimensions $S_d$ has come 
from rotation symmetry.
The wavelet coefficients 
\begin{equation}
\phi_a(b) = \int  a^{-d/2}\bar\psi\left(\frac{x-b}{a}\right)\phi(x)d^dx
\label{wtl2}
\end{equation}
represent the snapshot of the field $\phi(x)$ taken at the scale $a$ with 
the aperture function $\psi(x)$, and will be referred to as scale components 
of the field $\phi$. See e.g. \cite{Chui} for more details 
on wavelets.

The restriction imposed by the admissibility condition \eqref{adc} on 
the fiducial vector $\psi$ (the basic wavelet) is rather loose: 
only the finiteness of the integral $C_\psi$ given by \eqref{adcf} is 
required.  
This practically implies only that $\int \psi(x)dx =0$ and that 
$\psi(x)$ has compact support. For this reason the wavelet transform 
\eqref{wtl2}
can be considered as a microscopic slice of the function $\phi(x)$ taken 
at a position $b$ and resolution $a$ with ``aperture'' $\psi$. Of course,
 each particular aperture $\psi(x)$  has its own view, but the physical 
observables should be independent on it. In practical applications of 
WT  either of the derivatives of the Gaussian 
$\psi_n(x) = (-1)^n d^n/dx^n e^{-x^2/2}$ are often used, but for the 
purpose of this present paper only the admissibility condition is important but not 
the shape of $\psi(x)$. 

So, for the case of decomposition of a scalar field in $\R^d$ 
with respect to affine group \eqref{rag}, the inverse free field propagator matrix 
element is 
\begin{eqnarray*}
\bra a_1,b_1;\psi | D | a_2, b_2; \psi\ket = \int d^dx (a_1a_2)^{-\frac{d}{2}}
\bar\psi \left( \frac{x-b_1}{a_1} \right)  D \psi\left(\frac{x-b_2}{a_2}\right) \\
= \int \dk{k}{d} e^{ik(b_1-b_2)} (a_1a_2)^{\frac{d}{2}} \overline{\hat \psi(a_1 k)}
(k^2+m^2) \hat \psi(a_2 k) \\
\equiv \int \dk{k}{d} e^{ik(b_1-b_2)} D(a_1,a_2,k).
\end{eqnarray*} 
Assuming the homogeneity of the free field in space coordinate, i.e. that 
matrix elements depend only on the differences $(b_1\!-\!b_2)$ of the 
positions, 
but not the positions themselves, we can use  $(a,k)$ representation:
\begin{eqnarray}
\nonumber D(a_1,a_2,k) &=& a_1^{d/2} \overline {\hat \psi(a_1 k)} (k^2+m^2)
                 a_2^{d/2}\hat \psi(a_2 k) \\
D^{-1}(a_1,a_2,k) &=& a_1^{d/2} \overline {\hat \psi(a_1 k)} 
                  \left( \frac{1}{k^2+m^2} \right)
                 a_2^{d/2}\hat \psi(a_2 k)\\
\nonumber d\mu(a,k)       &=& \dk{k}{d}\da{a}{d}.
\end{eqnarray}
Thus, we have the same diagram technique as usually, but with extra 
``wavelet'' term $a^{d/2}\hat \psi(a k)$ term on each line and 
the integration over $d\mu(a,k)$ instead of $dk$.

Now, turning back to the coordinate representation \eqref{wtl2}, where 
$a$ is the resolution (``window width'') and recalling the power law 
dependence resulted from the Wilson expansion, we can define the 
$\phi^4$ model on affine group, with the coupling constant 
dependent on scale. The simplest case of the fourth power interaction 
of this type is
\begin{equation} 
V_{int}[\phi] =\int \frac{\lambda(a)}{4!} \phi^4_a(b) d\mu(a,b),
\quad  \lambda(a)\sim a^\nu.
\label{vint}
\end{equation}

The one-loop order contribution to the Green function $G_2$ 
in the theory with interaction \eqref{vint}
can be easily evaluated \cite{Alt2001} by integration over a scalar 
variable $z = ak$: 
\begin{equation}
\int \frac{a^\nu a^d |\hat\psi(ak)|^2 }{k^2+m^2}\dk{k}{d}\da{a}{d}  
= C_\psi^{(\nu)}  \int \dk{k}{d} \frac{k^{-\nu}}{k^2+m^2}, 
\end{equation} 
where $$C_\psi^{(\nu)}=\int |\hat\psi(z)|^2 z^{\nu-1}dz.$$ 
Therefore, there are no UV divergences for $\nu>d\!-\!2$. 

However, the positive values of $\nu$ mean that the interaction strengths at large 
scales and diminishes at small. This is a kind of asymptotically free 
theory that is hardly appropriate say to magnetic systems. What is required 
instead is a theory with the interaction vanishing outside a given 
domain of scales. Such models will be presented in the next section 
by means of scale-dependent stochastic quantization. 

\section{Stochastic quantization with wavelets}
The method of stochastic quantization 
first introduced by G.Parisi, and Y.Wu \cite{PW1981} consists in 
substitution of functional integration by averaging over certain random 
process. 
Let $S_E[\phi]$ be the action Euclidean field theory \eqref{fe} in $\R^d$.
Then, instead of direct calculation of the Green functions  
\eqref{gfm} from the generation functional \eqref{gf}, it is possible to 
introduce a fictitious 
``time'' variable $\tau$, make the quantum fields into stochastic fields 
$\phi(x) \to \phi(x,\tau),x\!\in\!\R^d,\tau\!\in\!\R $ and evaluate the 
moments  
$\langle \phi(x_1,\tau_1) \ldots \phi(x_m,\tau_m) \rangle$ by averaging over 
a random process $\phi(x,\tau,\cdot)$ governed by the Langevin equation 
\begin{equation}
\frac{\d\phi(x,\tau)}{\d\tau} +\frac{\sigma^2}{2} 
\frac{\delta S}{\delta\phi(x,\tau)} = \eta(x,\tau).
\label{le}
\end{equation}  
The gaussian random force $\eta$, that drives the Langevin equation 
\eqref{le},
has zero mean and is $\delta$-correlated in both the $\R^d$ coordinate and 
the fictitious time:
\begin{equation}
\langle \eta(x,\tau)\rangle=0, \quad 
\langle \eta(x,\tau)\eta(x',\tau') \rangle = \sigma^2 
\delta(x-x')\delta(\tau-\tau'). 
\label{gforce}
\end{equation}
The physical Green functions are obtained by taking the steady state limit
\begin{equation}
G(x_1,\ldots,x_m) = \lim_{\tau\to\infty} 
\langle \phi(x_1,\tau) \ldots \phi(x_m,\tau) \rangle
\label{ssl}
\end{equation}
either in the solution of the Fokker-Planck equation, or in the stochastic 
perturbation expansion given by stochastic generating functional 
\begin{equation}
W[J(x,\tau)] = \int \cD\phi \cD\eta 
\exp\left[{
\int d^dx d\tau \left( J\phi - \frac{\eta^2}{2\sigma^2}\right)
}\right] 
\delta \left[{
\frac{\d\phi}{\d\tau}+\frac{\sigma^2}{2}\frac{\delta S}{\delta\phi}-\eta
}\right].
\label{cfr}
\end{equation}
The stochastic quantization procedure has been considered as 
perspective candidate for the regularization of gauge theories, 
for it does not require gauge fixing. However 
$\delta$-correlated Gaussian random force in the Langevin equation still 
yields singularities in the perturbation theory. For this 
reason a number of modifications based
on the noise regularization
 $\eta(x,\tau) \to
\int dy R_{xy}(\d^2)\eta(y,\tau)$ have been proposed
\cite{BGZ1984,BHST1987,IP1988}. 
The introduction of a colored noise 
$\bra\eta(x,\tau)\eta(x',\tau')\ket = \delta(x-x')|\tau-\tau'|^n$ instead of 
$\delta$-correlated one can be also used to avoid UV divergences 
\cite{BGZ1984,Alfaro1985}. Other methods of regularization 
in stochastically quantized theories were also considered 
\cite{NY1983,CKS1988}.

In this paper we intend to revive the method of stochastic quantization by 
applying the continuous wavelet transform to both the fields $\phi(x,\tau)$ and 
the random force $\eta(x,\tau)$. It will be shown that ultra-violet divergences 
in so constructed perturbation expansion can be eliminated for some particular 
choice of the random force correlator taken in wavelet space. Our method is rather 
general and can be applied to any stochastic systems described by the Langevin 
equation. The idea of our method was proposed in \cite{AltDAN03} and consists in the 
following.

Instead 
of the usual space of the random functions  
$f(x,\cdot)\in(\Omega, \cA, P)$, where
$f(x,\omega)\in \lr$ for each given realization $\omega$ of the random process, we go  
to the multi-scale representation provided by the continuous wavelet 
transform \eqref{wtl2}: 
\begin{equation}
W_\psi(a,\vb,\cdot) = \int |a|^{-\frac{d}{2}}
\overline{\psi\left(\frac{\vx-\vb}{a}\right)}f(\vx,\cdot)d^dx.
\label{wtr}
\end{equation}
Since the structure of divergences and the localization of the solutions  
are determined by the spatial part of the random 
force correlator (see \eg \cite{KPZ1986}), the wavelet transform is 
performed only in the 
{\em spatial} argument $x\!\in\!\R^d$ of the dynamical variable $\phi(x,\tau)$, 
but not in its fictitious time argument.

The inverse wavelet transform 
\begin{equation}
f(\vx,\cdot) = C_\psi^{-1} \int |a|^{-\frac{d}{2}} 
\psi\left( \frac{\vx-\vb}{a}\right) W_\psi(a,\vb,\cdot) 
\frac{dad\vb}{a^{d+1}}
\label{fa}
\end{equation} 
reconstructs the common random process as a sum of its scale components, 
\ie projections onto different resolution spaces. 

The use of the scale components instead of 
the original stochastic process provides an extra analytical 
flexibility of the 
method:  there exist more than one set of random functions 
$W(a,\vb,\cdot)$ the images of which  have 
coinciding correlation functions in the space of $f(x,\cdot)$. 
It is easy to check that the random process generated by wavelet 
coefficients having in $(a,k)$ space the correlation function 
\begin{equation}
\bra \widehat W(a_1,k_1) \widehat W(a_2,k_2)\ket 
= C_\psi (2\pi)^{d} \delta^d(k_1+k_2)  
a_1^{d+1} \delta(a_1-a_2) D_0   
\label{w2}
\end{equation}
has the same correlation function as white noise has: 
$$
\begin{array}{lcl}
\bra f(x_1) f(x_2) \ket &=& D_0 \delta^d(x_1-x_2), \\
\bra {\hat f}(k_1) \hat f(k_2)\ket &=& (2\pi)^d D_0 \delta^d(k_1+k_2) \label{gn}\\
\nonumber \bra \widehat W_\psi(a_1,k_1) \widehat W_\psi(a_2,k_2)\ket &=&  
 (2\pi)^d D_0 \delta^d(k_1+k_2)
(a_1 a_2)^{d/2} \overline{\hat \psi(a_1 k_1) \hat\psi(a_2 k_2)}.   
\end{array}
$$
Therefore, starting from a given random process in the space 
of scale-dependent functions  $W(a,\vb,t)$, rather than in a  common 
space of square 
integrable functions $f(\vx,t)$, we can design a narrow band 
forcing with no contradictions
to other physical constraints. This can be done by   applying 
the requirement  $W(a,\vb,t)\!\to\!0$ for all $a$ outside a certain domain 
of scales $[a_{min},a_{max}]$.

Now let us turn to the stochastic quantization of the $\phi^3$ theory with 
the help of scale-dependent noise designed in the above described way.
The Euclidean action of the $\phi^3$ theory is 
\begin{equation}
S_E[\phi(x)]= \int d^d x \left[\frac{1}{2} (\d\phi)^2 + \frac{m^2}{2}\phi^2
+ \frac{\lambda}{3!}\phi^3 \right].
\label{pf3}
\end{equation}
Therefore the corresponding Langevin equation used for stochastic 
quantization is written as  
\begin{equation}
\frac{\d\phi(x,\tau)}{\d\tau} + \frac{D_0}{2}\left[ 
-\Delta \phi + m^2\phi + \frac{\lambda}{2!} \phi^2 \right] = \eta(x,\tau),
\label{le3}
\end{equation} 
where in common case the $\delta$-correlated random force is applied 
$$
\bra \eta(x,\tau) \eta(x',\tau')\ket = D_0 \delta^d(x-x')\delta(\tau-\tau').
$$ 
Following \cite{AltDAN03}, we perform continuous wavelet transform of the 
fields and forces in the spatial coordinate 
\begin{equation}
\phi(x)  = C_\psi^{-1} \int  \exp(\imath(\vk\vx-\omega\tau)) a^{\frac{d}{2}}
\hat\psi(a\vk) \hat \phi(a,k) \dk{k}{d+1} \da{a}{d},
\label{zt}
\end{equation}
using hereafter apparent $(d\!+\!1)$ dimensional notation $x=(\vx,\tau),k=(\vk,\omega)$. 
To generalize the scale-dependent force \eqref{w2} we introduce dependence on 
$k$ in the force correlator
\begin{equation}
\begin{array}{lcl}
\bra \widehat \eta(a_1,k_1) \widehat \eta(a_2,k_2)\ket 
&=&  C_\psi (2\pi)^{d+1} \delta^{d+1}(k_1+k_2)
a_1^{d+1} \delta(a_1-a_2)  2D(a_2,\vk_2),\\ \bra \hat\eta(a,k) \ket &=& 0,
\end{array}
\label{sfnc}
\end{equation}
which coincides with \eqref{w2} if $D(a,\vk)=const$, and is therefore capable 
of giving white noise in that limiting case. After substitution of 
\eqref{zt} into the Langevin equation \eqref{le3} we yield the stochastic 
integro-differential equation for stochastic fields $\hat\phi(a,k,\cdot)$
\begin{equation}
\begin{array}{l}
(-\imath\omega + \vk^2 + m^2)\hat\phi(a,k) =  \hat\eta(a,k)
-\frac{\lambda}{2} a^{\frac{d}{2}}  \overline{\hat\psi(a\vk)} 
C_\psi^{-2} \int (a_1 a_2)^\frac{d}{2}\\ 
\hat\psi(a_1\vk_1) \hat\psi(a_2(\vk-\vk_1))  
\hat \phi(a_1,k_1) \hat \phi(a_2,k-k_1) \dk{k_1}{d+1} \da{a_1}{d}\da{a_2}{d}.
\end{array}
\label{phi3w}
\end{equation} 

\begin{figure}[h]
\centering\includegraphics[width=8cm]{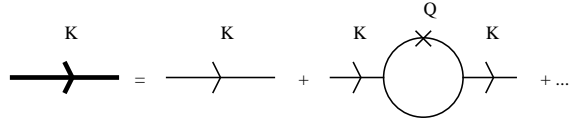}
\caption{Diagramm expansion of the stochastic Green function in $\phi^3$-model}
\label{gf:pic} 
\end{figure}

\begin{figure}[h]
\centering\includegraphics[width=8cm]{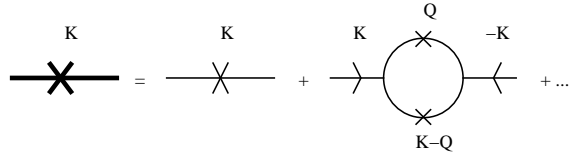}
\caption{Diagramm expansion of the stochastic pair correlation function in $\phi^3$ 
model}
\label{cf:pic} 
\end{figure}

Starting from the zero-th order approximation 
$\hat \phi_0(a,k) = G_0(k) \hat\eta(a, k)$ with the bare Green function  
$$G_0(k) = \frac{1}{-\imath\omega + \vk^2 + m^2}$$
and iterating the integral equation \eqref{phi3w}, 
in one loop approximation we get the correction to the stochastic 
Green function 
\begin{equation}
G(k) = G_0(k) + \lambda^2 G_0^2(k) \int \dk{q}{d+1} \Delta(\vq) 
|G_0(q)|^2 G_0(k-q) + O(\lambda^4),
\label{phi3G1}
\end{equation}
where $\Delta(\vk)$ is the scale averaged correlator \eqref{dak}:
\begin{equation}
\Delta(\vk) \equiv  C_\psi^{-1} \int \frac{da}{a} |\hat\psi(a\vk)|^2 2D(a,\vk) 
\label{dak}
\end{equation}  
In the same way all other momenta \eqref{gfm} can be evaluated. Thus the 
common stochastic 
diagram technique is reproduced with the scale-dependent random force
\eqref{sfnc} instead of the standard one.
The 1PI diagramms corresponding to the stochastic Green function decomposition 
\eqref{phi3G1} are shown in Fig.~\ref{gf:pic}.
Similarly, to stochastic Green function \eqref{phi3G1}, the one-loop 
contribution to the stochastic pair correlation function can be evaluated in 
$(a,k)$ space 
\begin{equation}
\bra \hat\phi(a_i,k)\hat\phi(a_f,-k) \ket \equiv C(a_i,a_f,k) 
= C_0(a_i,a_f,k) + \lambda^2 C_2(a_i,a_f,k)
+ O(\lambda^4),
\label{scf}
\end{equation}
where $$C_0(a_i,a_f,k)=\Delta(\vk)|G_0(k)|^2 (a_ia_f)^{d/2} 
\overline{\hat\psi(a_i\vk)\hat\psi(-a_f\vk)}.$$
The one-loop contribution to the pair correlator is 
\begin{eqnarray}
 C_2(a_i,a_f,k) &=& \frac{1}{2} |G_0(k)|^2 (a_i a_f)^{d/2} 
\overline{\hat \psi(a_i\vk)\hat \psi(-a_f\vk)} \label{phi3C1} \\ 
\nonumber & & \int \dk{q}{d+1} |G_0(q)|^2|G_0(k-q)|^2 
 \Delta(\vq) \Delta(\vk-\vq). 
\end{eqnarray}
The 1PI diagramms corresponding to the stochastic pair correlator decomposition 
\eqref{scf} are shown in Fig.~\ref{cf:pic}.
An important type of scale-dependent forcing is that acting on a single 
scale:
\begin{equation}
D(a,\vk)=\delta(a-a_0) D(\vk).
\label{sb}
\end{equation}
In some sense this mimics a field theory on a grid with a fixed mesh $a_0$.

As an example, let us consider the $\phi^3$ model with the single-scale 
force \eqref{sb} and the ``Mexican hat'' used as a basic wavelet  
\begin{equation}
\hat \psi(k) = (2\pi)^{d/2} (-\imath \vk)^2 \exp(-\vk^2/2), \quad 
C_\psi = (2\pi)^d.
\label{mh}
\end{equation}
This gives the effective force correlator 
\begin{equation}
\Delta(\vq) = 2 \frac{(a_0\vq)^4}{a_0} e^{-(a_0\vq)^2}D(\vq)
\label{efc2}.
\end{equation}
The loop integrals taken with this effective force 
correlator \eqref{efc2} can be easily seen to be free of 
ultra-violet divergences. The IR divergences also become milder 
because of the wavelet power factor $(a_0\vq)^4$. 

In fact, substituting \eqref{efc2} into 
expressions for one-loop contributions to the stochastic Green 
function and the correlation functions, \eqref{phi3G1} and \eqref{phi3C1}, 
respectively, we get 
\begin{eqnarray}
G_2(k)   &=& G_0^2(k) I_{G2}, \\
\nonumber I_{G2} &=& \int \dk{\vq}{d} \Delta(\vq) \int_{-\infty}^\infty 
\frac{d\Omega}{2\pi} \frac{1}{\Omega^2 + (\vq^2+m^2)^2}
\frac{1}{-\imath(\omega-\Omega) + (\vk-\vq)^2+m^2} \\ 
C_2(a_i,a_f,k) &=& \frac{1}{2} |G_0(k)|^2 (a_i a_f)^{d/2} 
\overline{\hat \psi(a_i\vk)\hat \psi(-a_f\vk)} I_{C2},\\ 
\nonumber I_{C2} &= & \int \dk{\vq}{d}  \Delta(\vq) \Delta(\vk-\vq)
\int_{-\infty}^\infty \frac{d\Omega}{2\pi} \frac{1}{\Omega^2 + (\vq^2+m^2)^2}
\frac{1}{(\omega-\Omega)^2 + [(\vk-\vq)^2+m^2]^2}.
\end{eqnarray}
For the case of single-scale forcing 
\eqref{efc2} the exponential factor in $\Delta(\vq)$ will suppress any 
power divergences comming from the Green functions. In fact, in a stationary 
limit ($\omega\!\to\!0$), after the integration over the frequency 
variable $\Omega$, we get 
\begin{eqnarray}
\lim_{\omega\to0} I_{G2} &=& 
\int \dk{\vq}{d} \Delta (\vq) \frac{1}{2(\vq^2+m^2)}
\cdot\frac{1}{\vq^2+(\vk-\vq)^2+2m^2} \\
\lim_{\omega\to0} I_{C2} &=& 
\int \dk{\vq}{d} \Delta (\vq) \frac{1}{2(\vq^2+m^2)((\vk-\vq)^2+m^2)}
\cdot\frac{1}{\vq^2+(\vk-\vq)^2+2m^2}.
\end{eqnarray}
An explicit calculation gives the frequency dependence:
\begin{eqnarray*}
I_{G2}&=& \frac{1}{2B}\frac{1}{(A-B)-\imath\omega} 
       +  \frac{1}{(\omega+\imath A)^2+B^2} \\
I_{C2}&=& \frac{1}{2A} \frac{1}{(\omega+\imath A)^2+B^2} 
       +  \frac{1}{2B} \frac{1}{(\omega-\imath B)^2+A^2} 
\end{eqnarray*}
where $$A = (\vk-\vq)^2 + m^2, \quad B = \vq^2 + m^2.$$ 

For the $\phi^4$ and higher polynomial models the above described method 
can be applied in a straightforward way. 
\section{Conclusion}
In this paper, that is extended version of \cite{AltG24}, we have presented 
a new approach to regularization of field theories based on wavelet 
decomposition of the fields.
The idea of using the affine group in quantum field theory is not 
new. It is in the basis of the renormalization group method the modern 
quantum field theory resides on. However, the direct use of the 
{\em representations} of affine group as a basis in the space of wave 
functions can provide a new perspectives in construction of a 
divergence free field theory. From the point of view of functional analysis  
this is 
an extension of the space of square integrable functions $\phi(x)$ to 
the space of scale-dependent functions $\phi_a(x)$, representing 
a snapshots of the field $\phi$ at a given resolutions $a$. The usage of 
functions depending on scale suggests that the interaction potential should 
be scale-dependent too. Such potentials having been de facto already in use 
in renormalization group technique, can be directly incorporated into quantum 
field theory from beginning. One of means to do it is the method of 
scale dependent stochastic quantization proposed in this paper. 
Although direct construction of the quantum field theory on the affine group 
presented in the beginning of this paper is also possible.  
\section*{Acknowledgement}
The author has benefited from discussions with Profs. S.Chaturvedi, 
H.H\"uffel, A.K.Kapoor, V.B.Priezzhev  and V.Srinivasan. This work 
is supported in part by Russian Foundation for Basic Research, 
project 03-01-00657.
discussions.

\end{document}